Title: Transient and quantal nucleation in solids: a statistical approach
Author: Mladen Georgiev (Institute of Solid State Physics, Bulgarian Academy of
 Sciences, 1784 Sofia, Bulgaria)
Comments: 6 pages including 3 figures, all pdf format
Subj-class: physics

In a recent arXiv preprint we proposed a statistical approach to the quantal effects in nucleation rate, addressed particularly to solid state physics. We now turn to transient nucleation rates incorporating either classical or quantal statistics. A similar method may be applied to tackle nucleation problems in anyon systems in solids.

## 1. Foreword

Nucleation has long been regarded as one of the most enigmatic phenomena in general [1], and in solids in particular [2]. Among the controversial features are the quantal effects, if any. Another feature is the transient effects regarded as something going deep into the heart of the matter [3]. Lately, we addressed the former feature by applying a statistical approach to the solutions of the eigenvalue problem of the scattering of a test particle by the surface and volume parts of a simple nucleus model [4]. The method, based on the reaction-rate theory, has been developed before to describe experimental temperature dependences of relaxation rates and times of impurities dissolved in crystalline solids. This yielded practical results to complement fundamental solutions [5].

We believe there is nothing wrong turning for a while attention from Volmer's master equation [6] in an attempt to develop a more pragmatic approach involving classical or quantal statistics [7]. Another feature that may be addressed statistically is the motion of anyons in crystalline planes controlled by fractional statistics [8].

The purpose of this paper is further advancing the statistical approach while applying it to specific experimental systems. The solid state provides a favorable environment where cluster evolution from embryos to critical nuclei and further may be followed by spectroscopic means [9], as done at the dawn of fast flash-light or laser technology.

## 2. Steady-state nucleation rate

The proposed statistical definition of a steady-state nucleation rate $\aleph(r)$ rests on the definition of a phonon-coupled chemical reaction rate $\Re(r)$ [4]. The latter is the total transition rate summed up of the transition probabilities (rates) at all vibronic energy levels $E_n$ across a radial double-well potential $E_{d-w}(r)$ [5]. Taken in a similar way as a radial potential, Gibbs' energy $\Delta G(r)$ draws a similar graphic dependence comprising an interwell barrier at $r_C$ between two metastable sites. The similarity suggests using the same math model to derive the associated component rates. On doing that, the so-defined nucleation rate $\aleph(r)$ scales with the reaction rate $\Re(r)$, especially at the lowest temperatures where the quantal effects predominate [4]. This makes harder attributing the observed rates to nucleation or reaction processes.

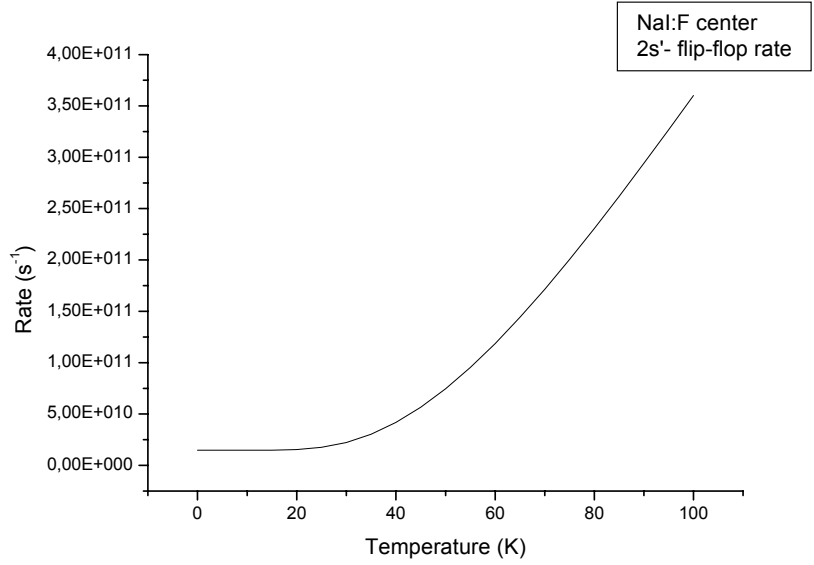

Figure 1: Reaction rate vs. temperature T in a flip-flop process in NaBr. Two temperature ranges are to be distinguished (from left to right): low T (0-20K) where the rate is insensitive to the temperature, intermediate T (20-50K) and high T (50-100K) where the rate is barrier controlled. In an earlier publication (Reference [4]) we presented arguments that the statistical nucleation rate should depend on the temperature in a similar fashion.

The steady-state rate reads (Boltzmann statistics) [4]:

$$\Re(r) = \nu (Z_A/Z_O) \sum_n W(E_n, r_C) \exp(-E_n / k_B T) \qquad (1)$$

where $\nu$ and $Z_A$ are the frequency and partition function of the reactive mode, $Z_O$ is the partition function of all the modes, reactive and accepting, $E_n$ are the quantized energies. $W(E_n)$ are the transition probabilities across the barrier, tunneling for $E_n < E_B$. For harmonic modes $Z_A/Z_O = 2\sinh(\frac{1}{2}h\nu/k_B T) = [\exp(\frac{1}{2}h\nu/k_B T) - \exp(-\frac{1}{2}h\nu/k_B T)]$ which when multiplied by $\exp(-\frac{1}{2}h\nu/k_B T)$ from the $E_0$ energy exponent eliminates the related first-term divergency eventually leading to a nonvanishing zero-point rate, and indeed it yields $\nu W(E_0)$ at $E_0 = \frac{1}{2}h\nu$. A typical rate dependence on temperature is drawn in Figure 1 attributing the essential ranges to fundamental steps.

### 3. Transient nucleation rate

#### 3.1. Descending trends

We use a simplistic model to make our point clear. Extensions are straightforward.

The rate defines a relaxation time $\tau(r) \propto \aleph(r)^{-1}$ which satisfies the linear rate equation

$$dN(r,t) / dt = -N(r,t) / \tau(r) \qquad (2)$$

which readily solves to give

$$N(r,t) = N(r_0) \exp(-t / \tau(r)) \tag{3}$$

It is said that within a time-interval of length ~ $4\tau(r)$ the nucleation is transient [3] or that it runs under transient conditions. From (2) and (3) we define an effective transient rate

$$\aleph(r,t)_{desc} \equiv -N(r_0)^{-1} dN(r,t) / dt = \exp(-t/\tau(r)) / \tau(r) = \aleph(r) \exp[-\aleph(r)t] \tag{4}$$

which tends to vanish as $t \to \infty$ though at $t = 0$, $\aleph(r,0)_{eff} = \aleph(r)$. We also get equivalently

$$\aleph(r,t)_{desc} = -(d/dt) \exp[-\aleph(r)t] \tag{4'}$$

The descending transient nucleation rate $\aleph(r,t)_{eff}$ from (4) is shown in Figure 2.

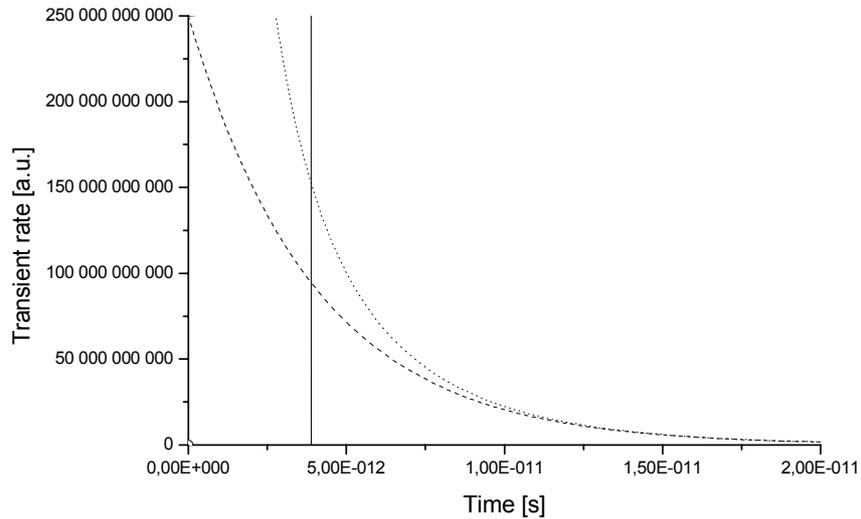

Figure 2: Proposed descending transient nucleation rate from equation (4) (dash) and ascending transient nucleation rate from equation (7) (dot). The zero-point steady-state rate is used at $2.5 \times 10^{11}$ $s^{-1}$. The vertical line at $3.5 \times 10^{-12}$ s is retained for convenience.

### 3.2. Ascending trends

The ascending case obtains as a time-independent generating rate $\Gamma(r)$ is superimposed onto a system of nuclei. We get the rate equation:

$$dN(r,t) / dt = \Gamma(r) - N(r,t) / \tau(r) \tag{5}$$

whose solution is

$$N(r,t) = \exp(-t / \tau) \int_0^t dt' \, \Gamma(r) \exp(+t' / \tau) = \Gamma(r) \tau(r) [1 - \exp(-t / \tau(r))] \tag{6}$$

As above, we define a transient nucleation rate

$$\aleph(r,t)_{asc} = N(r,t)^{-1} dN(r,t) / dt = [1/\tau(r)] \exp(-t/\tau(r)) / [1-\exp(-t/\tau(r))] =$$

$$\aleph(r) \exp(-\aleph(r)t) / [1-\exp(-\aleph(r)t)] \propto \aleph(r) \exp(-\aleph(r)t) \tag{7}$$

Again, this tends to vanish as $t \to \infty$ even though at $t = 0$, $\aleph(r,0)_{eff} = \aleph(r)$.

$$\aleph(r,t)_{asc} = (d/dt) \ln [1-\exp(-\aleph(r)t)] \tag{7'}$$

The ascending transient nucleation rate is also shown graphically in Figure 2.

In a real experiment, there will be both ascending and descending trends, so as the above analysis may be found useful for the proper interpretation of experimental data.

From equations (4) and (7) we see the ratio $\aleph(r,t)_{desc} / \aleph(r,t)_{asc} = [1-\exp(-\aleph(r)t)] < 1$ which asymmetry, however, tends to vanish at $t \to \infty$.

## 4. Statistics controlled nucleation rates

### 4.1. Boson pairing and nucleation

Bosons coagulate directly to form a superfluid state, since they obey Bose-Einstein (BE) statistics which allows them to occupy the energy ground state, all at a time. Superfluidity is, therefore, an extreme manifestation of quantum nucleation by bosons.

The BE distribution function is [7]

$$F_{BE}(E_n) = 1 / ( \exp[(E_n - \mu) / k_B T] - 1 ) \tag{8}$$

$\mu$ is the chemical potential of the bosonic system.

The particle-exchange phase $|\psi_1\psi_2\rangle = e^{i\theta} |\psi_2\psi_1\rangle$ is $e^{i\theta} = 1$ ($\theta = 0, 2\pi$) for bosons.

### 4.2. Fermion pairing and nucleation

Unlike bosons, fermions cannot occupy the energy ground state or any given quantum state including the spin more than one particle at a time. For this reason they may not condense to form a superfluid, unless by forming pairs beforehand which pairs behave like bosons. There are two ways for fermions to form bosonic pairs by way of: (a) momentum-space pairing (Cooper pairs) and (b) real-space pairing by forming fermionic quasimolecules. The theory of k-space pairing (*pairing theory*) is considered by many to be one of the utmost achievements of human thought.

The fermions obey Fermi-Dirac (FD) statistics whose distribution function reads [7]

$$F_{FD}(E_n) = 1 / ( \exp[(E_n - \mu) / k_B T] + 1 ) \tag{9}$$

where µ is the Fermi energy (chemical potential). At zero-point T = 0, $F_{FD}(E_n) = 0$ ($E_n > E_F$) while $F_{FD}(E_n) = 1$ ($E_n \leq E_F$).

The particle-exchange phase $|\psi_1\psi_2\rangle = e^{i\theta} |\psi_2\psi_1\rangle$ is $e^{i\theta} = -1$ ($\theta = \pi$) for fermions.

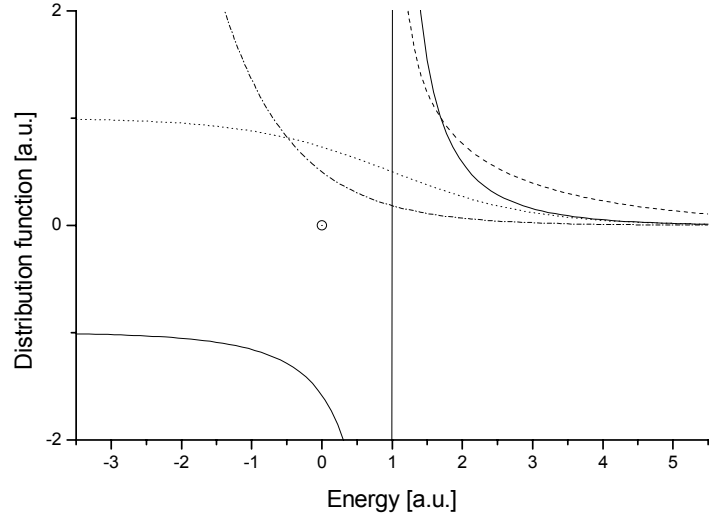

Figure 3: Graphic sketch of the energy distribution functions of statistics controlled nucleation: classic (BS) (dash-dot), quantal Fermi-Dirac (FD) (dot), quantal Bose-Einstein (BE) (solid), and quantal semion fractional statistics (FS) (dash).

### 4.3. Anyon pairing and nucleation

This is the case of fractional statistics intermediate between bosons ($\theta = 0, 2\pi$) and fermions ($\theta = \pi$). Anyons form on a plane or on a surface (sphere, torus) [10].

Perhaps the central position among anyons is occupied by the particles with phase constants $\theta = \pi/2$ and $\theta = 3\pi/2$ called semions, with phases $+i$ and $-i$, respectively [8]. Pairs of $+i$ semions form bosons and could give rise to superfluidity, so could triplets of $-i$ semions which form bosons as well [8]. We find the literature on anyons less than illuminating for the time being and so is our further reference-dependent discussion of this very interesting item.

Given that semion pairs behave like bosons we postulate a $+i$ semion distribution function of the form

$$F_{SI}(E_n) = 1 / \sqrt{(\exp[(E_n - \mu) / k_B T] - 1)} \tag{10}$$

### 4.4. Boltzmann statistics

To obtain a rate under each particular statistics, we insert equations (8), (9), and (10), into equation (1) to replace the Boltzmann factors

$$F_B(E_n) = (1/Z_O) \exp(-E_n/k_B T), \tag{11}$$

respectively. (Equation (11) represents the energy distribution function under Boltzmann's classical statistics.) By this manipulation we substitute a quantal statistics for the classical statistics. So far we have no ready-to-use practical solution to cover the general fractional statistics in lieu of equation (10).

The distribution functions for various statistics from equations (8) through (11) are sketched graphically in Figure 3.

## 5. Conclusion

We introduced a statistical approach to the transient nucleation rates by considering two examples of descending and ascending trends, respectively. It is believed these might be found useful for analyzing experimental data. In an independent study we revisited the quantal statistics, as done earlier [7]. This time we also included the presumed semion fractional statistics to open a dialogue with anyon researchers.

The heart of the matter of our statistical approach are the trans-barrier probabilities $W(E_n,r)$ as derived by a method due to John Bardeen [11]. So far we have dealt with two configurations involving parabolic and trigonometric barriers as applied to reaction rates and local rotations, respectively [4,5]. The dome-shaped nucleation barrier will be considered in greater detail in a subsequent publication.

## Acknowledgement

I am thankful to Professor D.S. Nenow (Bulgarian Academy of Sciences) for his comments.